\include{graphicsx}

\documentclass{aastex}
\slugcomment{Accepted for publication in ApJ Letters}

\shorttitle{Gl 569B: Multiple BD System}
\shortauthors{Kenworthy et al.}

\begin{document}


\title{Gliese 569B: A young multiple brown dwarf system?}


\author{Matthew Kenworthy$^1$, Karl-Heinz Hofmann$^2$, Laird Close$^1$,
Phil Hinz$^1$, Eric Mamajek$^1$, Dieter Schertl$^2$, Gerd Weigelt$^2$,
Roger Angel$^1$, Yuri Y. Balega$^3$, Joannah Hinz$^1$, George Rieke$^1$}

\email{mak@as.arizona.edu}

\affil{$^1$Steward Observatory, University of Arizona, Tucson, AZ 85721}
\affil{$^2$Max Planck Institut f\"{u}r Radioastronomie, D-53121 Bonn, Germany}
\affil{$^3$Special Astrophysical Observatory, Nizhnij Arkhyz,\\ 
Zelenchuk region, Karachai-Cherkesia, 357147 Russia}   



\begin{abstract} 

The nearby late M star Gliese 569B was recently found by adaptive optics
imaging to be a double with separation $\sim 1$ AU. To explore the
orbital motion and masses, we have undertaken a high resolution
($\sim0\farcs05$) astrometric study.  Images were obtained over 1.5
years with bispectrum speckle interferometry at the 6.5m MMT and 6m SAO
telescopes. Our data show motion corresponding to more than half the
orbital period, and constrain the total mass to be $> 0.115\,M_\odot$,
with a most probable value of $0.145\,M_\odot$.  Higher masses cannot be
excluded without more extended observations, but from statistical
analysis we find an 80\% probability that the total mass is less than
$0.21\,M_\odot$.

An infrared spectrum of the blended B double obtained with the MMT has
been modeled as a blend of two different spectral types, chosen to be
consistent with the measured $J$ and $K$ band brightness difference of a
factor $\sim2$.  The blended fit is not nearly as good as that to a pure
M8.5+ template.  Therefore we hypothesize that the brighter component
likely has two unresolved components with near equal masses, each the
same as the fainter component. 

If Gl 569B is a triple our dynamical limits suggest each component has a
mass of $50^{+23}_{-4}\,M_{jup}$.  We infer an age for the system of 300
Myr, from its kinematic motion which places its as a member of the Ursa
Major moving group. All the above parameters are consistent with the
latest DUSTY evolutiuon models for brown dwarfs.

\end{abstract}

\keywords{binaries: general --- stars: evolution --- stars: formation
--- stars: individual (Gl 569) --- stars: low-mass, brown dwarfs}

\section{Introduction}

Gliese 569 has been known for some time as a binary star 9.8 pc distant,
with two M stars at a projected separation of 50 AU.  The low mass
secondary (subsequently designated 569B) was found through infrared
imaging \citep{for88}.  Visible imaging and low resolution spectroscopy
of this component suggested it to be a M8.5 dwarf \citep{hen90} although
rather luminous for its age when compared to other cool field dwarfs
\citep{for88}.  In 1999, \cite{mar00} (hereafter M00) imaged the fainter
B component with the Keck II adaptive optics system and found it to be a
binary with projected separation of 1 AU and $\sim0.7$ magnitude
brightness difference.  A determination of the mass of the system from
orbital motion is thus possible and of particular interest, since it is
thought (from stellar activity indicators) to be young, between 0.1 and
1.0 Gyr (M00).  At this age the components of Gl 569B are good brown
dwarf candidates.  There has been little chance to test brown dwarf
models against objects of known mass from direct dynamic measurement,
although a lower mass limit was obtained by \cite{bas99} for the
spectroscopic brown dwarf binary PPl~15.

\section{Astrometric imaging by bispectrum speckle interferometry}

We imaged the Gl 569 system with the Bonn IR speckle camera on 2000 July
4 at the new 6.5m Multi-Mirror Telescope (MMT), 10 months after the
original discovery image (M00), and again on 2001 March 9 and 10 at the
Special Astrophysical Observatory (SAO) 6m telescope in Russia.  In both
cases reimaging optics were used so as to properly sample diffraction
limited speckles.  At the MMT the images were recorded in the $H$ and
$K$-band with pixel sizes of $18.70\pm0.19\,mas$ and
$24.68\pm0.25\,mas$, and at the 6m telescope images were recorded in
both the J and H bands, with $13.33\pm0.13\,mas$ and $
20.11\pm0.20\,mas$ respectively. The seeing was typically
1\farcs0---1\farcs5 over all nights.

\begin{figure}
\includegraphics[angle=0,width=\columnwidth]{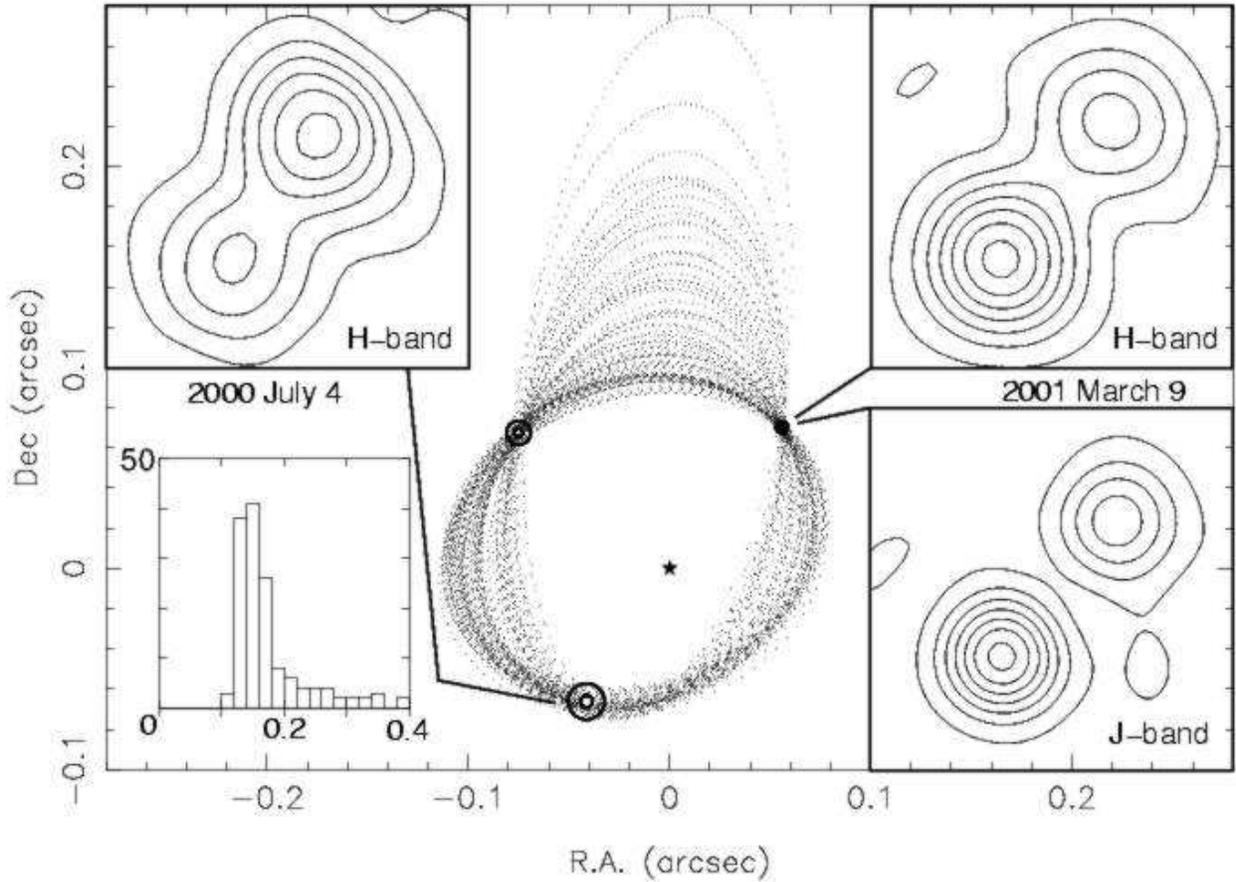} \caption{The
orbital solutions found which all pass within $2\sigma$ of each observed
epoch. The solid inner circles are $1\sigma$ error bars and the outer
circles are $3\sigma$. Ba is at the center of the frame with Bb orbiting
it. The MMT and SAO bispectrum speckle images (inset) are shown with the
highest contours at 95\% of the peak value and decreasing in steps of
10\%. The orbital solutions and the reconstructed images have the same
scale. North is up and East to the left for all images. The histogram in
the lower left corner shows the distribution of mass (in units of Solar
mass) for all accepted
orbital solutions.\label{fig1}} \end{figure}

\placefigure{fig1}

Diffraction-limited images of the triple system were reconstructed using
the bispectrum speckle interferometry method described in \cite{wei77},
\cite{loh83} and \cite{hof86}. The bispectrum of each frame consisted of
113 million elements and the object power spectrum was determined with
the speckle interferometry method, \citep{lab70}. The unresolved bright
point source Gl 569A of the wide ($5\farcs0$) binary served as a
reference star for the determination of the speckle transfer function
and the resulting images are diffraction-limited images (Figure
\ref{fig1}).  The pixel scale was derived from measurements of several
wide calibration binaries with well-known separation and position angle
(with separation error $\le$ 1\% and position angle error $\le 1\degr$).  

The position angles, radial separation and associated measurement errors
are given in Table \ref{tbl-1}.  Also listed for completeness is the
published discovery data from the Keck II telescope (M00).

\placetable{tbl-1}

\section{Orbital solution for Ba/Bb}

Solutions for the orbital elements of the close binary Ba/Bb were found
from the three epochs of observation, using classical astrometric
techniques \citep{ait64}.  Although the data points are well placed to
cover most of the orbit, a unique solution is not possible without more
extended observations.  Nevertheless, we are able to place analytically
a strong lower limit to the combined mass.  Orbits which exactly fit the
data are not possible for a combined mass of less than $0.136\,M_\odot$,
and for a wide range of assumed eccentricity (0.3 to 0.5) the mass lies
between 0.136 and 0.150 solar masses, with the corresponding periods
from $2.3-3.5$ yr and the semi-major axes lying between $0.93-1.25$ AU.

In order to explore the effect of measurement errors and the possibility
that we happened to catch a higher mass system with wider spacing at
higher inclination, we constructed a Monte Carlo model.  Many trial
binaries were constructed with uniform distribution in total mass,
ellipticity and period.  The epoch of periastron ($t$) was chosen
randomly within the range $0 < t < P$ with $P$ up to 10 yr and viewing
directions were modeled as from points uniformly distributed over a
sphere. The ephemeris was calculated and the calculated positions for
the three observed epochs compared with the observations. If the orbit
matched the data to within $2\sigma$ of each of the three observed data
points, the orbital elements of that orbit were noted, along with the
combined mass of the system.

The mass distribution found in this way is non-gaussian (see inset
Figure \ref{fig1}), with 80\% in the range $0.115-0.216\,M_\odot$ and
the remaining highest 20\% forming a high mass tail.  We conclude that
the combined mass for Gl 569B is $0.144^{+0.059}_{-0.010}\,M_\odot$ for
20\%---80\% limits, with a hard lower mass limit of $0.115\,M_\odot$,
consistent with the analytically derived orbital fits.  Our present
astrometry cannot yield the division of mass between the individual
components. 

\section{Spectral types and temperatures of Gl 569Ba and Gl
569Bb\label{specevidence}}

We obtained $J, H, K$ spectra of Gl 569A and the B component blend on 4
March 2001 1230 UT at the 6.5m MMT Observatory with the FSPEC IR
spectrograph \citep{wil93}.  FSPEC is a cryogenic long slit near-IR
($1-2.5\micron $) spectrograph which we used at the low $(R\sim700)$
resolution mode.  The spectra were taken and reduced with standard IR
beam-switching techniques. Terrestrial lines were removed by observing a
F9V star just before and after the Gl 569 science exposure, and at a
nearly identical airmass.  The F9 and the Gl 569B spectra were extracted
with standard IRAF routines and wavelength calibrated with the OH night
sky emission lines. No contamination from Gl 569A was observed. 

\begin{figure}
\includegraphics[angle=0,width=\columnwidth]{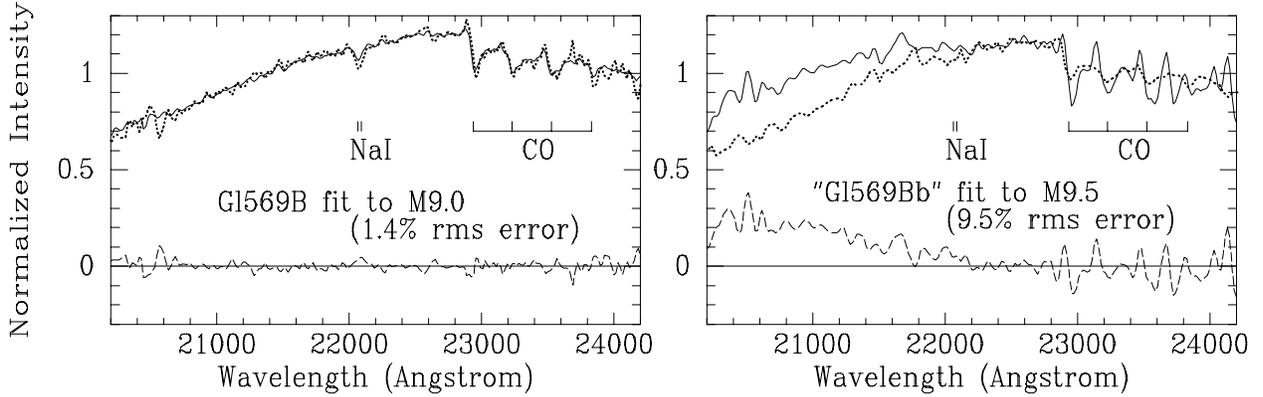}

\caption{The left-hand plot shows a fit of the observed GL569B $K$ band
spectrum (solid line) with our M9.0V template star (BRI1222-1221; dotted
line). The excellent match (residuals are plotted as the dashed line)
shows the blended light from Gl~569Ba and Bb appears consistent with
just a M9.0 spectrum. The plot on the right is similiar except that the
residual ``Gl~569Bb'' de-blended spectrum is fit to a M9.5.  Even though
this ``Gl~569Bb'' was our best deblended spectrum and M9.5 was the
closest spectral match in the $K$ band, it is shown to have a poor fit
with a large residual error. Hence, our deblending efforts of Gl~569B's
spectra into a hotter star and a cooler are much less successful than
just a pure M8.5-M9 spectrum.\label{fig2}}

\end{figure}

We have compared the observed K-band spectrum of Gl 569B (containing
both Ba and Bb) to late M star dwarf standards taken during the same run
and have modeled it as a sum of two spectra corresponding to the
observed difference in $K$ magnitude of 0.7. It appears that a
minimization of the residuals occurs when ``deblending'' Gl 569Ba by
65.5\% M8.0 and 34.5\% M9.5. In Figure \ref{fig2} we show the effective
spectrum of ``Gl 569Bb'' assuming that 65\% of Gl 569B's light is from
Gl 569Ba being a M8.0 star.  This ``Gl 569Bb'' spectrum appears in the K
band to be closest to a M9.5.  We also show a fit of ``Gl 569Bb'' to a
M9.5 template and show 9.4\% rms error in the fit. Even though this was
our best blended fit, there are large residuals and poor fit to the CO
features. No mix of a cooler and hotter star weighted by $\Delta K =
0.7\,$mag works well.

\placefigure{fig2}
\placetable{eqwidth}

However, much to our suprise, we found that the IR $(1-2.5\micron)$
spectrum from Gl 569B was very well fit by a pure M9 template spectrum.
The rms residual from just fitting Gl 569B to a M9 spectrum yielded a
fit of only 1.4\% rms error and excellent fits to the CO features.

We examined our J spectra and measured the equivalent widths for both
the \ion{K}{1} doublets and the steam feature at $1.34\micron $ (see
Table \ref{eqwidth}). Based on these measurements and the best spectral
type indices of \cite{rei01} we find that Gl~569B is easily and
consistently classed as a M8.5-M9 (M8.5+) with a $T_{eff}$ of
$2150\pm75\,$K from the template scale of \cite{leg01}. These
temperature errors do not take into account the $\sim 300$K systematic
offsets between different models, $\pm75\,$K is simply a relative
temperature error. This choice of temperature scale is optimal since the
same Ames dusty atmospheric features and opacities were used in the full
DUSTY tracks \citep{cha00,cha01} and so any systematic offsets in the
stellar temperatures and the DUSTY model temperatures are minimized with
this choice of temperature scale. Moreover, as discussed in \cite{leg01}
this temperature scale has a relative accuracy ($\pm75\,$K) for the
M7-L3 SpT considered here where dust opacity plays a large role.

Again, when we attempt a separation as two components of different
brightness, the J band residual spectrum ``Gl 569Bb'' extracted from the
Gl 569B spectrum appears poorly fit by any template. However, we can
classify this ``Gl 569Bb'' somewhere in the M9.5-L3 range with a
$T_{eff}$ of $1985\pm120\,$K. The large error in this $T_{eff}$ again
suggests that this extracted ``Gl 569Bb'' spectrum is not physical and
so does not match any spectral type well. Therefore it appears both Bb
and Ba are best fit by an M8.5+ spectrum for both the bright and faint
components.

\section{Age of the Gliese 569 System}

Since age is critical to expected luminosity for such late stars, we
have examined the kinematic evidence, with a view to obtaining a more
accurate estimate.  We calculate the heliocentric UVW velocity for
Gl~569A using the proper motion and parallax from Hipparcos (HIP 72444),
the radial velocity of \cite{mar87} ($v_{r}$ = --7.17 $\pm$ 0.28 km
s$^{-1}$) and employ a galactic motion vector algorithm (J. Skuljan
2000, private communication). We find a UVW vector of (+7.8, +3.2,
--13.3) km s$^{-1}$, with uncertainties of (0.2, 0.1, 0.3) km s$^{-1}$.
Comparing this vector with \cite{sod90} scatter plots of the UVW motions
of nearby active dwarfs, we noticed that it is very close to that of the
Ursa Major (UMa) moving group (+12.6, +2.1, --8.0; \cite{sod93}), with
Gl 569 within 7.2 km s$^{-1}$ of Soderblom's space motion for the UMa
group. 

The age of the UMa nucleus is $\simeq$0.3 Gyr \citep{sod93}, however
there is a somewhat larger spread in age when one examines the
early-type stellar content of the UMa moving group on larger scales
\citep{asi99,che97}. From the correlation between the young age inferred
from stellar activity indicators, and the kinematic similarity between
Gl 569 and the UMa moving group, we suggest that it is a member of the
moving group with an age of $0.3\pm0.1$ Gyr.  We also note that the
slightly subsolar metallicity of Gl 569A ([M/H] = --0.15; \cite{zbo98})
is similar to the value for the Ursa Major nucleus stars ([Fe/H] =
--0.08\,$\pm$\,0.09; \cite{sod93}).

\section{Discussion}

We have seen in Section \ref{specevidence} that the blended spectrum of
Ba/Bb matches that of a single M8.5+ star with much smaller residuals
than a blend of an M8.0 and M9.5.  We therefore postulate that all the
light from Gl 569B (containing both Ba and Bb) is from an M8.5+ spectral
type. Moreover, we have independently found that the $\Delta J
- \Delta K$ differential colors of the Ba and Bb component are
  $-0.10\pm0.14\,$mag, in close agreement with the value of $0.0\pm0.14$
observed by M00. Therefore, we see no evidence of Bb being any redder
than Ba while appearing only half as bright! As Table \ref{eqwidth}
points out, the expected $\Delta J - \Delta K$ color for a binary
composed of a M8.0 and a M9.5 is $\Delta J
- \Delta K = (J-K)_{M9.5}-(J-K)_{M8.0} = 0.30$ mag. Since 0.3 mag is
  inconsistent (at $3\sigma$) with the $\Delta J - \Delta K =
-0.10\pm0.14$ observed it is difficult to understand how Gl 569B can be
composed of 2 stars of different spectral types. 

It appears that both Gl~569Ba and Gl~569Bb have the same temperature.
However, since Gl 569Ba is $1.9\pm0.2$ times as bright as Gl~569Bb, the
most logical explanation for this overluminosity is that Gl~569Ba is
{\slshape itself} a binary star (as first suggested by M00).  Moreover,
the lack of any blended spectral features cooler than M9 in the Gl~569B
spectra argue that the Gl 569Ba binary is likely composed of 2 stars
both close to M8.5-M9.0 in spectral type. Hence, we conclude that Gl
569Ba is most likely a close ($\le$ 0.1 AU) binary with nearly equal
magnitude components.  Therefore the Gl 569B system becomes a
hierarchical triple with Gl 569Bb orbiting around a binary Gl~569Baa and
Gl~569Bab. All 3 of these stars should have nearly identical M8.5+
spectral types and therefore very similar masses.

We now examine the models models of \cite{cha00,cha01} and see how
treating Ba/Bb as a double system compares to a triple system model.  In
order to do this, we adopt the photometry of the combined Gl~569B system
from \cite{for88}, who measured $K = 9.56\pm0.1$.  The individual
absolute magnitudes follow from our measured brightness ratios given in
Table \ref{tbl-1} and the parallax $d=101.91\pm1.67\,mas$ from Hipparcos
\citep{per97}. We take the values to be $M_K(Ba)=10.02\pm0.12$ mag and
$M_K(Bb)=10.72\pm0.12$ mag.

\begin{figure}
\includegraphics[angle=270,width=\columnwidth]{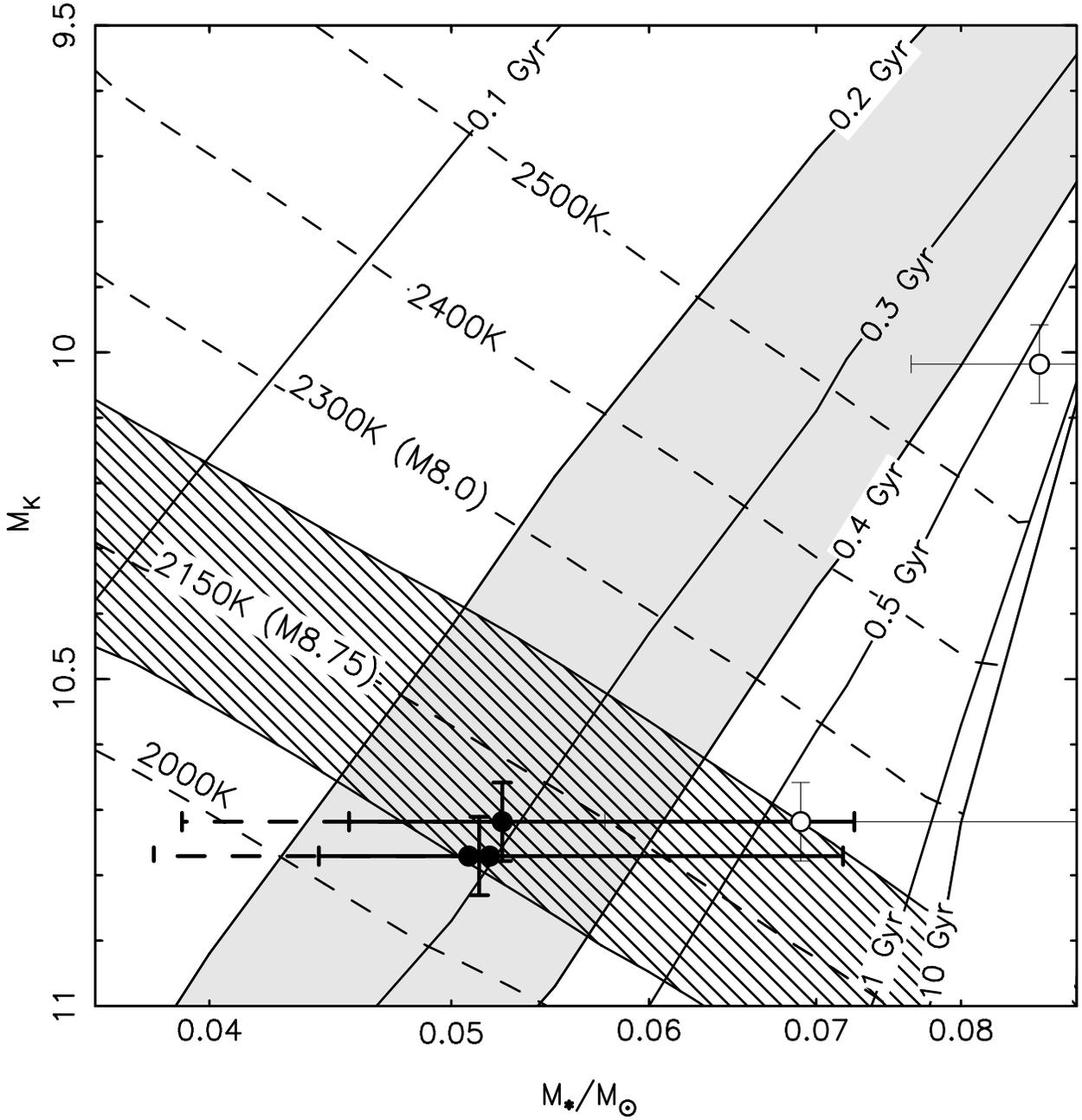}
\caption{Mass-$M_{K}$ diagram for Gl569Ba/Bb for the cases where Ba is
either a single or a double star. The shaded isochrones represent the
estimated age range for the UMa stream. The hollow circles represent the
positions of Ba and Bb if Ba is a single star. The solid circles
represent Ba and Bb if Ba is treated as an equal mass binary. In both
cases, mass error bars are for $1\sigma$ confidence limits and the hard
lower mass limit is represented as a dotted extension to the mass error
bars. The isotherms have SpT associated with them according to models
from \cite{cha00,leg01}.\label{fig3}}

\end{figure}

\placefigure{fig3}

Figure \ref{fig3} shows Ba/Bb considered both as a double system (hollow
circled points) and as a triple system with Ba composed of two equal
mass objects (filled circles). For the binary case, it is clear that
although Bb gives marginal agreement to its uncertain spectral type of
M9-L3, it is Ba that stands out as an object much more luminous than
spectral fitting and typing to an M8.0 would suggest. However,
considering B to be a triple system results in all three components of
nearly equal mass (the two Ba components are $0.049\,M_\odot$ each and Bb
is $0.057\,M_\odot$), a SpT consistent with an M8.5+, a $\Delta J - \Delta
K  \sim 0$ (as observed), and a model age in good agreement with a
kinematically derived age of $0.3\pm0.1\,$Gyr.

The astrometric and spectroscopic results presented here suggest that Gl
569B is a young heirarchical triple brown dwarf system with three nearly
equal components of $\sim50^{+23}_{-4}\,M_{jup}$ each, and a dynamically
constrained lower mass sum of $M = 0.115\,M_\odot$.  The work reported
here must therefore be regarded as simply a step to understanding what
promises to be a key brown dwarf system.  Continued high accuracy
astrometric measurements, as represented by our third epoch measure
($\pm1\,mas$), should yield an accurate and unambiguous total mass for
the system. Furthermore, by careful calibration of plate scale, accurate
measurement of the two stars individual motions should be possible, so
individual masses can be derived with no recourse to theoretical models,
and high resolution spectroscopy is required to see if Ba is indeed a
spectroscopic binary.

\acknowledgements

Some of the observations reported here were obtained at the MMT
Observatory, a joint facility of the University of Arizona and the
Smithsonian Institution.  We thank Isabelle Baraffe for supplying us with
theoretical models computed by France Allard, Isabelle Baraffe, Gilles
Chabrier and Peter Hauschildt, and Mike Meyer for helpful discussions.
MAK acknowledges support by the AFOSR under F49620-00-1-0294.

%% Tables should be submitted one per page, so put a \clearpage before

%% deluxetable environment provided by the AASTeX package or the LaTeX
%% table environment.  Use of deluxetable is preferred.
%%

%% Three table samples follow, two marked up in the deluxetable environment,
%% one marked up as a LaTeX table.

%% In this first example, note that the \tabletypesize{}
%% command has been used to reduce the font size of the table.
%% Note also that the \label command needs to be placed 

\clearpage
\begin{deluxetable}{llcccccc}
\tabletypesize{\scriptsize}
\tablecaption{Measured properties of the Gliese 569~B System\label{tbl-1}}
\tablewidth{0pt}
\tablehead{
\colhead{Epoch} &
\colhead{Telescope} &
\colhead{PA} &
\colhead{Separation} &
\colhead{} &
\colhead{} &
\colhead{} &
\colhead{Resolution} \\
\colhead{} & 
\colhead{} & 
\colhead{(deg)} &
\colhead{(mas)} &
\colhead{$\Delta J$} &
\colhead{$\Delta H$} &
\colhead{$\Delta K$} &
\colhead{(mas)} 
}
\startdata
1999.654 & Keck 10m\tablenotemark{a} & \phn48$\pm$2 &    101.0$\pm$2.0  & 0.5$\pm$0.2 & 0.5$\pm$0.1 & 0.5$\pm$0.1 & 50 \\
2000.501 & MMT 6.5m &    148$\pm$3 & \phn78.0$\pm$3.0  & \nodata & 0.7$\pm$0.1 & 0.7$\pm$0.1 & 53 \\
2001.186 & SAO 6.0m &    321$\pm$1 & \phn89.6$\pm$1.0  & \nodata & 0.9$\pm$0.1 & \nodata & 57 \\
2001.189 & SAO 6.0m &    320$\pm$1 & \phn89.9$\pm$1.0  & 0.6$\pm$0.1 & \nodata & \nodata & 53 \\
\enddata
\tablenotetext{a}{Results from \cite{mar00} included for comparison}
\end{deluxetable}

\clearpage
\begin{deluxetable}{lllllllllc}
\rotate
\tabletypesize{\scriptsize}
\tablecaption{Measured equivalent widths for Gl 569B\label{eqwidth}}
\tablewidth{0pt}
\tablehead{
\colhead{Star} &
\colhead{\ion{K}{1}} &
\colhead{\ion{K}{1}} &
\colhead{\ion{K}{1}} &
\colhead{\ion{K}{1}} &
\colhead{H$_2$O} &
\colhead{K-Band SpT} &
\colhead{Adopted SpT} &
\colhead{Adopted $T_{eff}\tablenotemark{a}$} &
\colhead{Typical $(J-K)\tablenotemark{b}$} \\
\colhead{} &
\colhead{11690\AA} &
\colhead{11770\AA} &
\colhead{12440\AA} &
\colhead{12530\AA} &
\colhead{$1.34/1.29\micron$} &
\colhead{$2.05-2.40\micron$} &
\colhead{} &
\colhead{(K)} &
\colhead{for Adopted SpT}
}
\startdata
GL569B      & \phn5.48 (M8) &\phn7.52 (M9)  &$\phn9.93$ (M9) &\phn9.44 (M8.5) & 0.76 (M9) & M8.5+ & M8.5+   & $2150\pm75\phn$  & 1.20\\ 
``GL569Bb'' & 10.13 (L3)    &10.42 (L3)     &13.00 (L3)   &12.80 (L3)      & 0.68 (L1) & M9.5  & M9.5-L3 & $1985\pm120$ & 1.45\\  
M8 template\tablenotemark{c} & \phn5.29 (M8) &\phn7.12 (M8.5)&$\phn8.72$ (M8.0) &\phn7.40 (M7.5) & 0.81 (M7) & M8.0  & M8.0    & $2225\pm75\phn$  & 1.05\\
\enddata

\tablenotetext{a}{temperature scale from \cite{leg01}}
\tablenotetext{b}{typical colors from \cite{rei01}}
\tablenotetext{c}{Our M8 template star was 2MASSW J1444171+300214}
\tablecomments{all spectral types fit to the SpT indices of
\cite{rei01}}

\end{deluxetable}

\clearpage

\clearpage

\clearpage

%% The following command ends your manuscript. LaTeX will ignore any text
\end{document}